\documentclass[%
aip,
 jmp,%
 amsmath,amssymb,
 reprint,%
]{revtex4-1}

\usepackage{graphicx}
\usepackage{dcolumn}
\usepackage{bm}
\usepackage{color}

\begin{document}

\preprint{AIP/123-QED}

\title{Nonmonotonic fracture behavior of polymer nanocomposites}% Force line breaks with \\

\author{J. G. de Castro}
 \affiliation{Van der Waals-Zeeman Institute, Institute of Physics, University of Amsterdam, 1098 XH Amsterdam, the Netherlands}
\author{R. Zargar}
 \email{Zargar.Rojman@gmail.com.}
 \affiliation{Van der Waals-Zeeman Institute, Institute of Physics, University of Amsterdam, 1098 XH Amsterdam, the Netherlands}
\author{M. Habibi}
 \affiliation{Van der Waals-Zeeman Institute, Institute of Physics, University of Amsterdam, 1098 XH Amsterdam, the Netherlands}
\author{S. H. Varol}
 \affiliation{Department of Molecular Spectroscopy, Max Planck Institute for Polymer Research, Ackermannweg 10, 55128 Mainz, Germany}
\author{S. H. Parekh}
 \affiliation{Max Planck Institute for Polymer Research, Department of Molecular spectroscopy, Ackermannweg 10, 55128 Mainz, Germany}
\author{B. Hosseinkhani}
 \affiliation{SKF Engineering \& Research Center, PO Box 2350, 3430DT Nieuwegein, The Netherlands}
\author{M. Adda-Bedia}
 \affiliation{Laboratoire de Physique Statistique de l'Ecole Normale Superieure, 24 rue Lhomond, 75231 Paris cedex 05, France}
\author{D. Bonn}
 \affiliation{Van der Waals-Zeeman Institute, Institute of Physics, University of Amsterdam, 1098 XH Amsterdam, the Netherlands}

\date{\today}

\begin{abstract}
Polymer composite materials are widely used for their exceptional mechanical properties, notably their ability to resist large deformations. Here we examine the failure stress and strain of rubbers reinforced by varying amounts of nano-sized silica particles. We find that small amounts of silica increase the fracture stress and strain, but too much filler makes the material become brittle and consequently fracture happens at small deformations. We thus find that as a function of the amount of filler there is an optimum in the breaking resistance at intermediate filler concentrations. We use a modified Griffith theory to establish a direct relation between the material properties and the fracture behavior that agrees with the experiment.
\end{abstract}



\maketitle

\section{Introduction}

The mechanisms of failure of materials under stress are of paramount importance because of their use in a wide variety of applications \cite{Freund,Kanninen,Cadwell,Rivilin,Griffith}. Brittle materials usually break at very small deformations, typically on the order of a percent or less. In addition, their fracture behavior is difficult to reproduce, since much of the fracture properties are due to the existence of defects in the material \cite{Marder}. In practical situations, large deformations are not uncommon; it is exactly for this reason that composite polymer based materials are abundantly used. Rubbers are the prototypical polymeric materials that typically fracture at very high deformations, often at deformations exceeding $100\%$.  Usually, such rubber materials are reinforced by adding nano-sized filler particles to increase their modulus and toughness. These composite materials are widely employed; however the main challenge remains to predict the fracture behavior of these rubbers \cite{Mars,Mullins,Gent59,Gent87,Gent90,Diaz,Cam,Chen,Zhang,Donnet} as a function of their material properties.
\par In this Letter, we study the fracture behavior of rubbers filled with silica nanoparticles, which is a common way to improve the mechanical properties of the rubbers \cite{Chen,Botti,Tiwari}. We determine the stress and deformation at which the material fails for different concentrations and types (sizes) of filler particles. Our main finding is that the stress and the deformation at failure are non-monotonic: they pass through a maximum at intermediate filler concentrations. To rationalize these findings, we first examine the standard Griffith theory for brittle fracture, which uses an energy balance between the elastic energy gained upon propagation of a fracture and the surface energy lost by creating additional interfacial area \cite{Griffith,Peresson}. From this, we conclude that the energy barrier for the spontaneous nucleation of an initial fracture is so large that thermally-driven fluctuations are much too weak to cause spontaneous breaking at a given stress. We subsequently extend the Griffith theory using an Erying-type model that incorporates a stress-induced crossing of the energy barrier for crack formation. This allows to relate the stress at break of a filled rubber to the volume fraction of filler material based only on the fracture energy and modulus of the material, both of which can be measured separately.

\section{Experiments}

The materials used for this work are composites of Acrylonitrile Butadiene Rubber (NBR) filled with silica nanoparticles that were prepared at SKF Elgin, USA. In brief, commercial NBR of molar mass $M_w=2.5 \times 10^5 g/mol$, glass transition temperature $T_g= -36^{\circ} {\rm C}$ and mass density  $\rho=0.96 g/cm^3$ is used; the fillers are precipitated silica with three different primary particle sizes: $28$, $20$ and $15 nm$, which we name Silica1, Silica2 and Silica3 respectively. The amount of silica loaded in the NBR matrix is between $5$ and $90 phr$ (parts per hundred parts of rubber by weight) which covers a range of filler volume fractions from $1.59\%$ to $22.46\%$.
%%%%%%%%%%%%%%%%%%%%%%%%%%%%%%%%%
\begin{figure*}
\centering
\includegraphics[width=1.80\columnwidth]{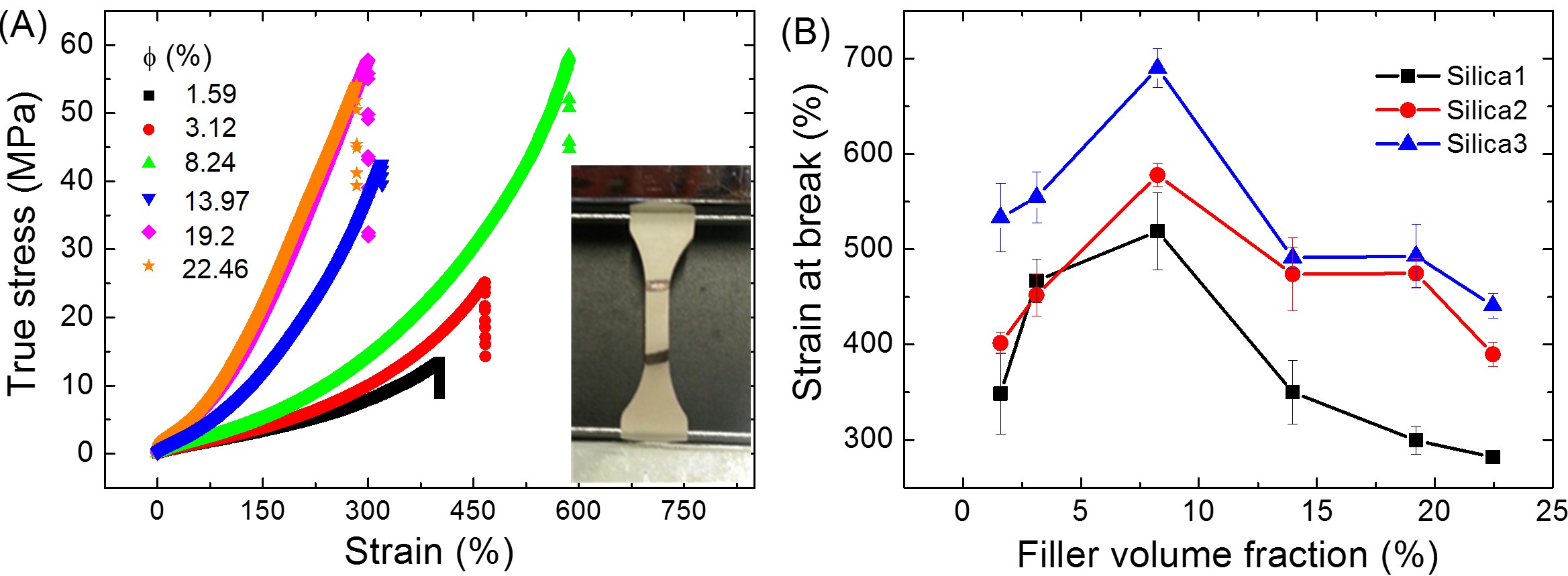}
\caption{A) True stress vs. strain for compounds of Silica1 type with different filler concentration. A sharp decrease of the stress at the end shows the break point. Inset shows the dumbbell-shaped specimens with the gauge marks used in the first series of experiments.  B) critical strain at break vs. volume fraction for all types of silica. Black color is for Silica1, red for Silica2 and blue for Silica3.}
\label{Fig1}
\end{figure*}
%%%%%%%%%%%%%%%%%%%%%%%%%%%%%%%%%
%The formulation of the rubber compounds is shown in Table \ref{Table1}.
%%%%%%%%%%%%%%%%%%%%%%%%%%%%%%%%%%
%\begin{center}
%\begin{table}[h]\footnotesize
%  \caption{Compounding formulation of the NBR rubber.}
%    \begin{tabular}{| l | r |}
%    \hline
%    Ingredient & phr (parts per hundred rubber) \\
%    \hline
%    NBR	& 100 \\ \hline
%    Stearic acid & 1 \\ \hline
%    ZnO & 9 \\ \hline
%    Rubber activator & 2.5 \\ \hline
%    Sulfur & 1.2 \\ \hline
%    Accelerator for curing agent & 2.5 \\ \hline
%    Silica1 (28 nm) & 5-90  \\ \hline
%    Silica2 (20 nm) & 5-90  \\ \hline
%    Silica3 (15 nm) & 5-90  \\ \hline
%    \end{tabular}
% \label{Table1}
% \end{table}
%\end{center}
%%%%%%%%%%%%%%%%%%%%%%%%%%%%%%%%%%
%\par Oscillatory measurements are carried out with an Anton Paar Physica MCR 300 rheometer mounted with a plate-plate geometry. The samples are cylinders of $5mm$ diameter and about $2mm$ thick and are glued onto the rheometer plates with cyanoacrylate glue (Loctite 480). The measurements are done at room temperature and frequency $1Hz$.
\par The mechanical testing of compounds is performed on a Zwick extensometer. Two series of tests are performed. In the first series, tensile test on dumbbell-shaped specimens are carried out according to the standard ASTM D412-98a (see Fig. \ref{Fig1}A inset). The grip separation speed is fixed on $500 \pm 50 mm/min$, with a preload of $1 N$. We test three to five samples for each compound at room temperature. With this experiment we measure force-displacement curves. Assuming that the material is incompressible, this can be transformed into true stress and deformation (Fig. \ref{Fig1}A). The true stress  can be defined as: $\sigma = FL/A_0L_0$, in which $F$ indicates the force, $L_0$ and $L$ are respectively the initial and actual distance between the gauge marks (see Fig. \ref{Fig1}A inset), and $A_0$ indicates the cross sectional area of the undeformed specimen. The strain is measured as $\gamma(\%)=\frac{L-L_0}{L_0} \times 100$.

\section{Results and Discussion}

%%%%%%%%%%%%%%%%%%%%%%%%%%%%%%%%%
\begin{figure}
\includegraphics[width=0.90\columnwidth]{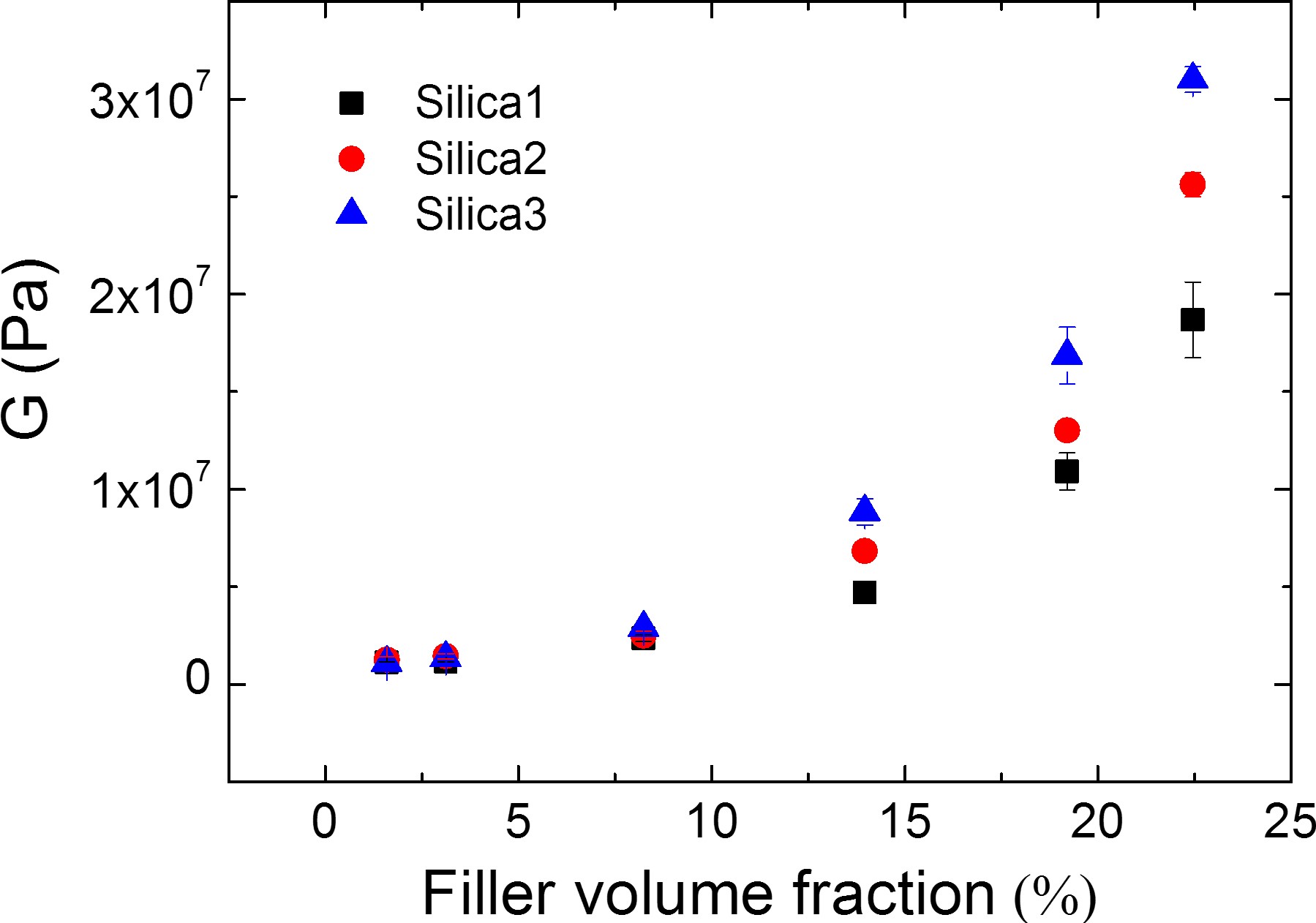}
\caption{Elastic modulus G versus filler volume fraction  for the three types of silica. Black color is Silica1, red for Silica2 and blue for Silica3. The Silica particle size decreases from Silica1 to Silica3.}
\label{Fig2}
\end{figure}
%%%%%%%%%%%%%%%%%%%%%%%%%%%%%%%%%
The curves of  Figure \ref{Fig1}A show a sharp drop of the true stress at the end that indicates the breaking of the sample. It follows that both the true stress and the deformation at break go through a maximum at intermediate filler concentration. For the other two silica types tested, the results are very similar to those shown in Figure \ref{Fig1}A. In all samples, the strain at break shows a peak around a volume fraction of $8\%$ filler (Fig. \ref{Fig1}B). At this filler concentration, the rubber shows the largest resistance against breaking, and since the overall deformation is larger, the stress at break is also larger. When the concentration of filler is higher than $8\%$, the material becomes more brittle and breaks at smaller deformations again; we also observe that the stress at break does not increase significantly after this concentration.
%%%%%%%%%%%%%%%%%%%%%%%%%%%%%%%%%
\begin{figure*}
\centering
\includegraphics[width=1.80\columnwidth]{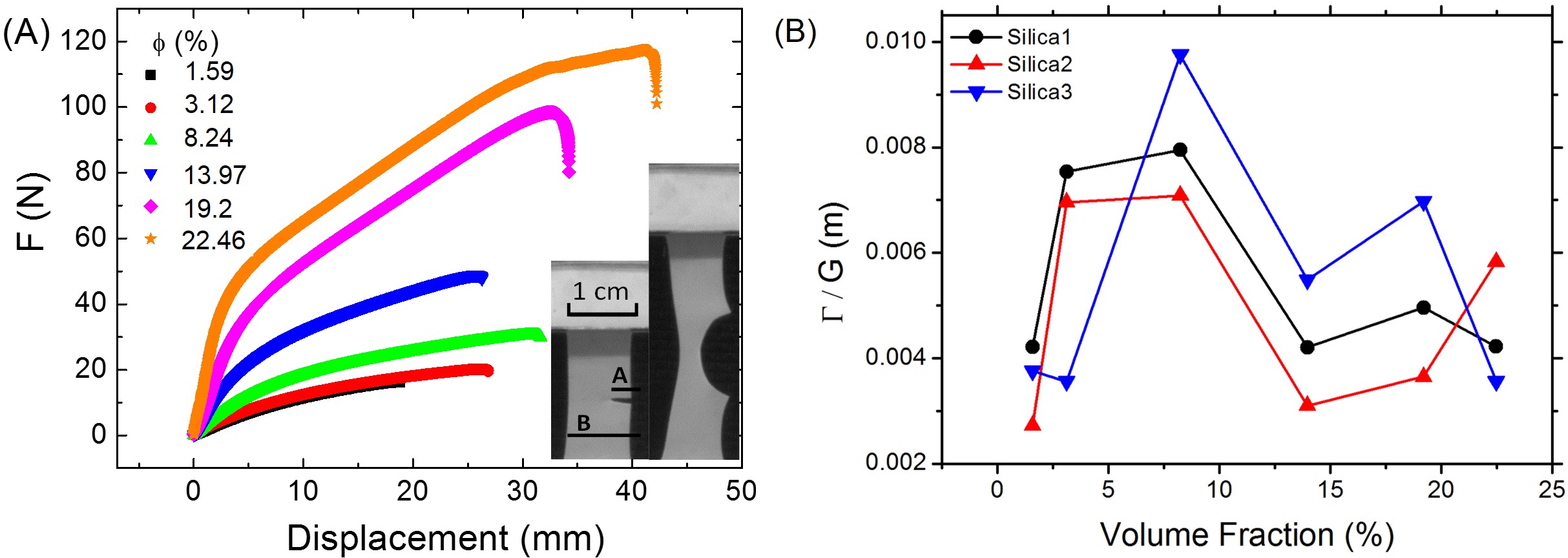}
\caption{A) Force vs. displacement for different filler concentrations in the second series of experiments where composites had an initial notch. Inset shows the fracture propagation process for NBR loaded with Silica1: the initial gauge length and the grip separation speed were respectively fixed on $2.5 cm$ and $10 mm/min$. B) Fracture energy scaled with the modulus versus the volume fraction of fillers for different types of silica.}
\label{Fig3}
\end{figure*}
%%%%%%%%%%%%%%%%%%%%%%%%%%%%%%%%%
\par To explain this behavior, we first characterize our materials. Measuring the Young's modulus at the relatively fast deformation rates of the tensile test is not accurate; we therefore determine the linear (visco-)elastic properties using standard rheology experiment. For our incompressible system, the Poisson ratio $\nu=0.5$, and the shear modulus is related to Young's modulus as $E=2G(1+\nu)=3G$. We find that while the measured shear modulus increases with increasing the filler concentration for all filler types, it increases more significantly for fillers with smaller particle size (Fig. \ref{Fig2}).
\par We subsequently determine the fracture energy $\Gamma$, required to create a fracture plane. The fracture energy $\Gamma$ includes not only the energy necessary to break the bonds at the crack tip, but also the energy dissipated in the vicinity of the crack tip during crack propagation \cite{Mullins,Peresson}. To determine $\Gamma$, we perform a second series of tensile tests on notched rectangular specimens with a width of $1 cm$ and thickness of $2 mm$. The samples were notched in the center, the depth of the notches being $2 mm$ (see Fig. \ref{Fig3}A inset).
\par From these experiments, we determine the fracture energy by calculating the work required to break the sample into two pieces and dividing that work by the created surface area. Figure \ref{Fig3}A shows the applied force $F$ on the system as a function of the displacement $\lambda$. The area under each curve gives the total work done on the samples up to their breakage. Assuming that all the work is used for creation of new surfaces, the fracture energy is obtained as:
\begin{equation}
\Gamma = \frac{\int_0^{\lambda_{max}}Fd\lambda}{2A_0}.
\label{FractureEnergy}
\end{equation}
\par What sets the force scale in these experiments is of course the elastic modulus; to scale out the trivial dependence of $\Gamma$ on the modulus, we scale the fracture energy with respect to the measured shear modulus for each sample.  Figure \ref{Fig3}B shows the scaled fracture energy for three filler types and different volume fractions. We find that, similar to the stress and the deformation at break, the scaled fracture energy shows a maximum at volume fraction about $8\%$ (see Fig. \ref{Fig1} and Fig. \ref{Fig3}B), meaning that here the samples are hardest to break, i.e., fail at the largest deformation. The nonmonotonic behavior of the fracture energy has been previously observed for nanosilica-epoxy resins \cite{Chen} however, we are the first to establish a theoretical framework to quantitatively explain this nonmonotonic behavior.
\par The question is now whether characterizing the bulk elastic properties and the fracture energy is sufficient to account notably for the non-monotonic fracture behavior (Fig. \ref{Fig1}). Classically, the energy barrier for the spontaneous formation (nucleation) of a crack is due to Griffith \cite{Griffith}: the energy barrier results from a competition between the cost in fracture (surface) energy and the gain in elastic (volume) energy for the formation of the initial crack \cite{Kanninen,Cadwell,Daniel,Noushine,Lawn}. In two dimensions, the surface energy cost $E_s$ of creating the crack depends linearly on the crack length $l$ and is given by $E_s\sim 2\Gamma l$ where $\Gamma$ is the fracture energy, and the elastic energy gain $E_v$ is quadratic in $l$ according to $E_v\sim 2\sigma^2 l^2/3G$, where $\sigma$ is the applied stress. The activation energy then follows from extremalization, i.e, finding the maximum of the total energy: $E_{barr-2D}\sim 3\Gamma^2 G /\sigma^2$. This equation shows a power-law dependence of $E_{barr}$ on $\sigma$, and hence the force, and has been confirmed for the fracture of two-dimensional crystals \cite{Pauchard}. The extension to the three-dimensional case gives \cite{Pomeau}:
\begin{equation}
U = -\frac{\sigma^2}{6G}(\frac{4}{3}\pi l^3)+2\Gamma \pi l^2.
\label{TotalEnenrgy}
\end{equation}
Its extremum corresponds to the energy barrier, occurs for $l_{crit}=\frac{6\Gamma G}{\sigma^2}$ and is
\begin{equation}
E_{barr-3D} = \frac{24\pi \Gamma^3 G^2}{\sigma^4}.
\label{BarrierEnenrgy}
\end{equation}
For the spontaneous nucleation of a crack in the system, thermal fluctuations should overcome this energy barrier, leading to a probability of fracture $P_{fracture}\sim \exp (\frac{-E_{barr}}{k_BT})$, where $T$ is the absolute temperature and $k_B$ is the Boltzmann constant \cite{Daniel}. Once overcome, a crack starts to propagate.
%%%%%%%%%%%%%%%%%%%%%%%%%%%%%%%%%
\begin{figure}
\includegraphics[width=0.90\columnwidth]{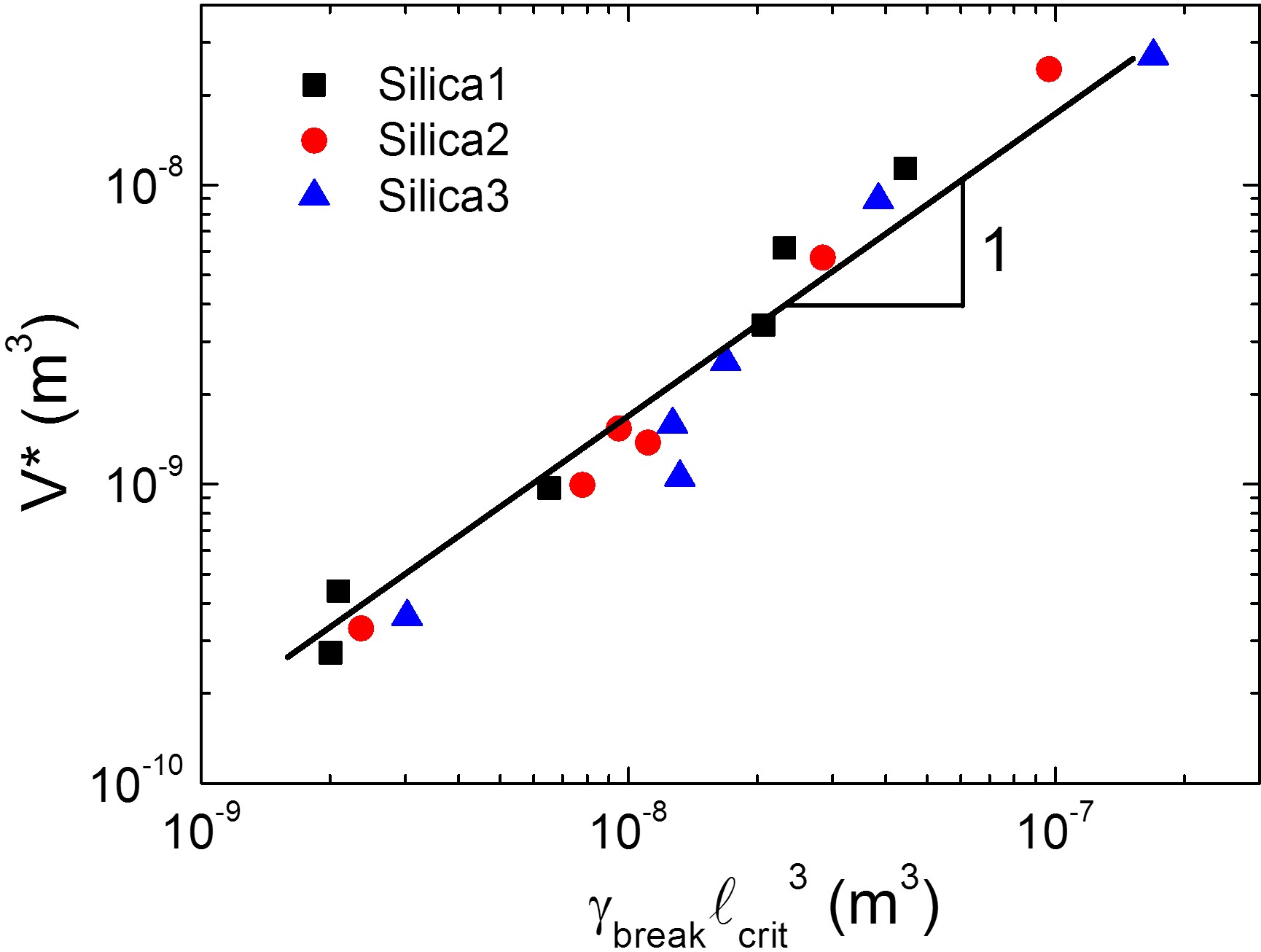}
\caption{Linear dependence between the activation volume $V^\ast$ and $\gamma_{break}l_{crit}^3$. Black color is for Silica1, red for Silica2 and blue for Silica3.}
\label{Fig4}
\end{figure}
%%%%%%%%%%%%%%%%%%%%%%%%%%%%%%%%%
We note that the above relations are obtained regardless of the filler particles. Hence, they are valid not only for silica particles but for any nano-sized hard particles.
\par Putting in typical numbers from the experiments, one immediately sees that $E_{barr}\gg k_BT$, and so a spontaneous, thermally activated, fracture \cite{Vanel} is not feasible in our system. This is because the elastomers dissipate enormous amounts of energy without breaking, which increases the fracture energy considerably and makes the energy barrier high compared to $k_BT$. The observation that the thermal energy alone is not sufficient to overcome the energy barrier is common for polymeric materials \cite{Atkins}, which has led to the consensus that for these systems the stress-induced crossing of the energy barrier may become important. This has led to a number of Eyring-type models that take into account the lowering of the energy barrier due to the applied stress \cite{Lee,Eyring,Richeton}. In its simplest form, the probability becomes:
\begin{equation}
P_{fracture}\sim \exp [\frac{-E_{act}+\sigma V^\ast}{k_BT}].
\label{FractureProbability}
\end{equation}
Zhurkov \cite{Zhurkov} provided a detailed comparison between the predictions of this model and the rate- and temperature dependent fracture and found that a wide range of polymeric materials follows this prediction.
\par In Eq.\ref{FractureProbability}, $V^\ast$ is the activation volume, which is often used as an adjustable parameter; if this is allowed, most experiments can be fit by the model. In our case, we define the activation volume $V^\ast \sim \gamma_{break}l_{crit}^3$ in which $\gamma_{break}$ is the strain at break that is experimentally measured (see Fig. \ref{Fig1}B). This definition is in fact necessary for consistency; since both $E_{act}$ (or $E_{barr-3D}$) and $\sigma V^\ast$ are much larger than the thermal energy, fracture will happen when $E_{act}\simeq\sigma V^\ast$. Putting in the expressions above for the energy barrier and the activation volume then leads to the familiar expression $\sigma_{break} \simeq G_{break}\gamma_{break}$, where $G_{break}$ is the slope of true stress versus strain very close to the breaking point (see Fig. \ref{Fig1}A).
\par To verify that the activation volume is indeed proportional to $\gamma_{break}l_{crit}^3$, we divided $E_{act}$ (calculated using Eqs. \ref{FractureEnergy} and \ref{BarrierEnenrgy}) by the experimental values of the stress at break (from Fig. \ref{Fig1}A) and compared these values to $\gamma_{break}l_{crit}^3$ (where $l_{crit}$ was calculated from Griffith's theory). Figure \ref{Fig4} shows the linear dependence of $V^\ast$ on $\gamma_{break}l_{crit}^3$.
\par
\begin{figure}
\includegraphics[width=0.90\columnwidth]{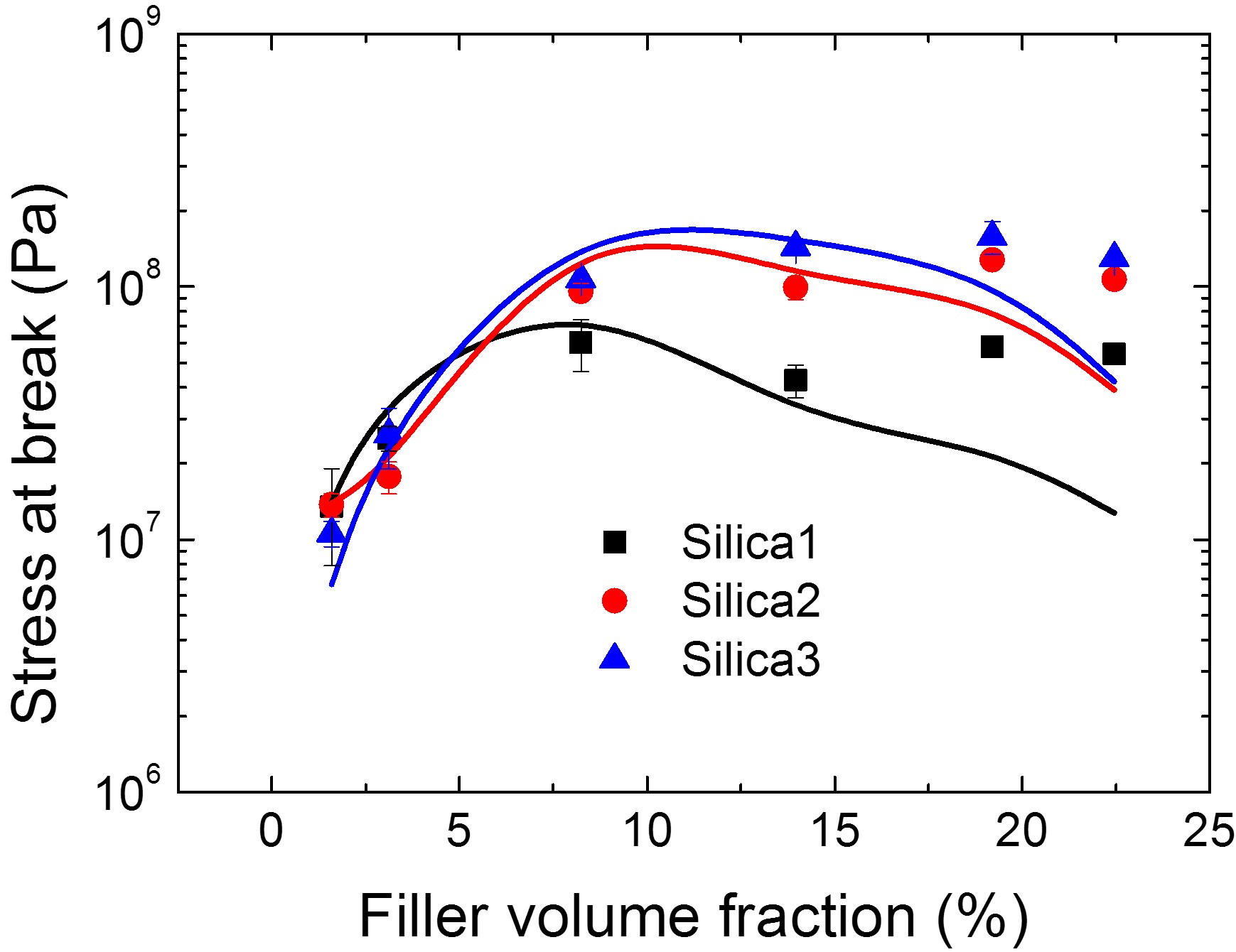}
\caption{Stress at break as a function of filler volume fraction: symbols are the experimental results and lines are the theoretical prediction using Eyring model and Griffith theory. Black color is for Silica1, red for Silica2 and blue for Silica3.}
\label{Fig5}
\end{figure}

 Having determined the activation volume in this way, we can obtain the breaking stress in our experiments. Figure \ref{Fig5} compares the calculated stress at break obtained from $\sigma = \frac{E_{act}}{\gamma_{break}l_{crit}^3}$ with the experimental results shown in Figure \ref{Fig1}A, using the experimentally determined values for $G$ and $\Gamma$. The experimental (symbols) and theoretical (lines) results are in very reasonable agreement, and reproduce the non-monotonic behavior of the stress at break. The theoretical values of the stress at break above $14\%$ volume fraction are somewhat lower than the experimental results; this is likely related to the large plastic deformation observed on those samples.
\par Note that we perform the same measurements for two different strain rates. We find that while the failure stress does not change with the strain rate, the failure strain increases with decreasing the strain rate.
%%%%%%%%%%%%%%%%%%%%%%%%%%%%%%%%%
%%%%%%%%%%%%%%%%%%%%%%%%%%%%%%%%%

\section{Summary}

In summary, we have experimentally established a direct relation between the material properties of our composite materials and the very non-linear problem of crack initiation that determines resistance to breaking.  For the composite materials considered here, we find that there exists an optimum amount of filler particles for which the filled rubbers show a maximum resistance against the applied stress and deformation and thus, are hardest to break. Using the adaptation of the Eyring model to the standard theory for fracture, we can explain how the non-monotonic fracture behavior is due to a subtle interplay between the bulk elastic energy gain and the surface fracture energy cost as a function of the filler concentration. These results should be relevant to filled polymeric systems in general, that all show a transition between the visco-elastic behavior of the polymer matrix without fillers and a more brittle behavior of the much harder composite material.

\section{Acknowledgments}

The authors thank the Stichting voor Fundamenteel Onderzoek der Materie (FOM), SKF, and Michelin for the financial support of the present research work.

\end{document}